\documentclass[
reprint,
amsmath,amssymb,
aps,
prl
]{revtex4-2}

\usepackage{hyperref}
\usepackage[utf8]{inputenc}

\hypersetup{colorlinks=true,
            linkcolor=black,
            citecolor=black,
            urlcolor=blue}
\usepackage{amssymb}
\usepackage{mathbbol}
\usepackage{courier}
\usepackage{dsfont}

\usepackage{physics}

\usepackage{graphicx}
\usepackage{dcolumn}
\usepackage{bm}

\begin{document}

\title{Quantum Desynchronization of Limit Cycles}

\author{Hans Christiansen}
\author{Jens Paaske}
\affiliation{Center for Quantum Devices, Niels Bohr Institute, University of Copenhagen, 2100 Copenhagen, Denmark}

\begin{abstract}
It is well known from classical physics that weakly coupled self-sustained oscillators may spontaneously lock their phases. Just like classical synchronization is known to break down due to noise induced phase slips, we show here how the synchronization of continuous variable quantum systems breaks down by proliferation of quantum phase slips. Within a Keldysh path integral formulation of limit cycles, we analyze the phase dynamics and show how, in spite of strong phase correlations, quantum phase slips degrade the actual phase locking. This approach also allows us to address non-Markovian effects on the synchronization of limit cycles, which we illustrate explicitly for oscillators coupled via a voltage biased double quantum dot. 
\end{abstract}

\maketitle

\textit{Introduction.} Driven-dissipative nonlinear oscillators can sustain stable oscillations whose amplitude and frequency are set by the dynamics, rather than by initial conditions. These are known as limit-cycle oscillators~\cite{Pikovsky2001Oct}. When two limit cycle oscillators interact and their respective frequencies are sufficiently close, they tend to synchronize their motion, meaning that their relative phase locks at a constant value. Inside an {\it Arnold tongue} where their frequency detuning is smaller than the interaction strength,  this phase locking is stable under weak perturbations. Numerous examples of synchronization of limit cycles can be found in nature, and provides, for example, the all-important stability of numerous biological functions under noisy conditions~\cite{LEFRANC2016507,WINFREE196715,Tabareau2010Jan}.  

Noise smoothens out the Arnold tongue and the desynchronization transition from locked to free-running relative phase turns into a crossover rather than a sharp transition. Physically, this happens because the relative phase starts slipping by sudden full revolutions of $2\pi$ due to the noise. These \textit{phase slips} cause diffusion of the relative phase which inhibits synchronization in the strict sense and renders the degree of synchronization a question of timescales~\cite{STRATONOVICH1965269, freund2003,Pikovsky2001Oct}.  


Improving the quality of synchronization by lowering the thermal fluctuations, one eventually reaches the lower bound set by the Heisenberg uncertainty principle. A number of recent theoretical and experimental works have addressed the potential synchronization of limit-cycle oscillators in the presence of quantum fluctuations, using a number of different observables as indicators of such {\it quantum synchronization} (cf. Refs.~\onlinecite{mari2013, walter_quantum_2014, walter_quantum_2015, liu2025observationsynchronizationquantumvan, nadolny2025quantumlimitcyclessynchronization} and references therein). Quantum phase slips play a prominent role in superconducting devices~\cite{Lau2001Nov, Mooij2006Mar, Pop2010Aug, hriscu_quantum_2013, Hohe2025Feb}, for which phase locking or slippage has important implications for the electronic properties of the system. Nevertheless, with the exception of Ref.~\onlinecite{Hassler2026May}, phase diffusion resulting from quantum phase slips has, to the best of our knowledge, not been employed to gauge the quality of quantum synchronization in the broader context of limit-cycle oscillators. Here we investigate the stability of limit cycles and their synchronization under the influence of quantum fluctuations.

First, we revisit the quantum Stuart-Landau equation~\footnote{This equation is often referred to as the quantum van der Pol equation, but we choose here to follow Ref.~\onlinecite{chia_relaxation_2020} and reserve that term to the more general equation which has no $U(1)$-symmetry and exhibits two characteristic time scales characteristic of a relaxation oscillator}, which has been defined in terms of a Lindblad master equation (LME) for a harmonic oscillator with jump operators describing single-photon gain and two-photon loss~\cite{walter_quantum_2015,lee_quantum_2013}. We demonstrate how quantum phase slips degrade synchronization, even in the presence of pronounced phase correlations. Next, we analyze a microscopic model for a non-Markovian resonant gain medium responsible for the synchronization of two oscillators and find that the synchronization frequency itself is entrained towards the resonance frequency of the environment.

\textit{Keldysh Action.} We begin from a model of a quantum harmonic oscillator with resonance frequency $\omega_0$, and self-energies appearing due to a microscopic environment $\Pi^{R,A,K}(\omega)$, described by a coherent state Keldysh path integral with an action given by~\cite{Kamenev2023Jan}
\begin{equation}
    S_2=\int\frac{\dd \omega}{2\pi}  \mqty( \bar\phi^{\rm cl} \\ \bar\phi^{\rm q} )_\omega^T\!\! \mqty( 0 & P^A_\omega  \\  P^R_\omega & P^K_\omega )
    \mqty( \phi^{\rm cl} \\ \phi^{\rm q} )_\omega 
\end{equation}
where $P^{R(A)}_{\omega}=\omega-\omega_0-\Pi^{R(A)}_{\omega}$ and $P^{K}_{\omega}=-\Pi^{K}_{\omega}$. The retarded Green function corresponding to this action is given as $D^{R}_{\omega}=(\omega-\omega_0-\Pi^R_{\omega})^{-1}$. If the environment serves as a gain medium, ${\rm Im}[\Pi_\omega^{R}]>0$ at $\omega \approx \omega_0$, its poles will lie in the upper half plane. This signifies an instability at the quadratic level, which we stabilize by adding the time-local quartic nonlinearity
\begin{equation}\label{eq:s4}
     S_4 = -\int \dd t  \mqty(\bar\phi^{\rm cl}\bar\phi^{\rm cl} \\ \bar\phi^{\rm cl}\bar\phi^{\rm q} \\ \bar\phi^{\rm q}\bar\phi^{q})_t^T
     \mqty( 0 & \Lambda_1 & \Lambda_2 \\ \Lambda_1^* & \Lambda_5 & \Lambda_3 \\ -\Lambda_2^* & \Lambda_3^* & \Lambda_4) 
     \mqty(\phi^{\rm cl}\phi^{\rm cl} \\ \phi^{\rm cl}\phi^{\rm q} \\ \phi^{\rm q}\phi^{\rm q})_t .
\end{equation}
This action maps directly to a quantum Stuart-Landau LME (cf. Supplementary Material (SM)~\footnote{See Supplemental Material at http: for details of the results presented in the main text.}) with jump operators $L_1 = \sqrt{\gamma_1}\hat{a}^\dag$ and $L_2=\sqrt{\gamma_2}\hat{a}^2$,  when the self energies are given as $\Pi^R =(\Pi^A)^*= i\gamma_1/2$ and $\Pi^K=-i\gamma_1$, $\Lambda_1=\Lambda_3^* = i\gamma_2/2$, $\Lambda_5=-2i\gamma_2$, and $\Lambda_2=\Lambda_4=0$~\cite{cottet_theory_2020,hermansen_synchronizing_2026,thompson_field_2023}. 

Anticipating limit-cycle solutions to the saddle-point equations with finite phase space radius $r>0$ and angular velocity $\nu$, it is convenient to transform the fields as
\begin{equation}
\mqty(\phi^{\rm cl}\\\phi^{\rm q}) = \mathrm{e}^{-i\theta - i\nu t}\mqty(r(1+\eta) \\  \chi/r) ,
\end{equation}
where $\theta$ and $\eta$ are real fluctuation fields, and $\chi$ is the rescaled complex quantum field. In terms of these fields the quadratic part of the action reads 
\begin{align}
    S_2[\theta,\eta,\bar\chi,\chi]&=\!\int\!\!\dd t \left[\mqty( \eta \\ \bar\chi)_t^{T}\!\!\mqty( 0 & P^A_{\nu + i D_t} \\  P^R_{\nu + i D_t}& P^K_{\nu+i D_t}/r^2 )
    \mqty(\eta \\ \chi )_t \notag\right.\\
    &\hspace{5mm}\left.+P^A_{\nu+\dot\theta}\chi_t + \bar\chi_t P^R_{\nu+\dot\theta} \right],
\end{align}
with $D_t=\partial_t-i\dot{\theta}$, and where $r$ and $\nu$ are determined by the stationary saddle-point equation
\begin{equation}\label{eq:MFequation}
0=\frac{\delta S}{\delta \bar\chi}= P^R_\nu-\Lambda_1^*r^2,
\end{equation}
together with its complex conjugate, $\delta S/\delta\chi=0$. A non-trivial solution, requires $r^2=P_\nu^R/\Lambda_1^*>0$, which determines the value of $\nu$~\cite{Sieberer_2016}. To analyze the fluctuations, we expand the quartic terms to second order in $\eta$ and $\chi/r$ to obtain 
\begin{align}\label{eq:s4}
S_4=-\!\!\int\!\! dt\,\qty[\Lambda_5 \bar\chi\chi + r^2(1+3\eta) (\Lambda_1^*\bar\chi+\Lambda_1\chi)],
\end{align}
omitting for simplicity the dephasing term, $\Lambda_2$~\cite{cottet_theory_2020}. 
Together with $S_2$, this results in the following effective action for the fluctuations
\begin{align}\label{eq:effectiveaction}
S_{\rm eff}&=\!\!\int\!\!\dd t \left[\mqty( \bar\eta \\ \bar\chi)^{T}\!\!
    \mqty(0 & P^A_{\rm eff}\\
          P^R_{\rm eff}& P^K_{\rm eff} )
    \mqty( \eta \\ \chi ) \right.\nonumber\\
&\left.\hspace{20pt}+\bar\chi \qty(P^R_{\nu+\dot\theta}-P^R_{\nu}) + (P^A_{\nu+\dot\theta}-P^A_\nu)\chi
    \right],
\end{align}
with $P^{R,A}_{\rm eff} = P^{R,A}_{\nu + i D_t}-3 P^{R,A}_{\nu}$ and 
$P^K_{\rm eff}=r^{-2}P^K_{\nu+iD_t}-\Lambda_5$.

\textit{Stuart-Landau Limit Cycle.} Taking $\Pi^{\alpha}$ and $\Lambda_{i}$ corresponding to the quantum Stuart-Landau equation, the saddle point values become $\nu=\omega_0$ and $r = \sqrt{\gamma_1/\gamma_2}$, and the effective action reduces to
\begin{align}
    &S_{\rm SL} =\!\!\int\!\!\dd t\qty[(\bar\chi+\chi)\dot\theta 
    +  \!\mqty( \bar\eta  \\ \bar \chi )^T\!\!\mqty( & i\partial_t  -i\gamma_1 \\  i\partial_t +i\gamma_1 & 3i\gamma_2 )
    \!\mqty( \eta \\ \chi)],
\end{align}
where we omitted also terms like $\bar\eta \dot\theta \chi$, assuming the diffusive dynamics of $\theta$ to be slow, as confirmed by the solution below. 
Eliminating the term to second order in $\chi$ by a complex Hubbard-Stratonovich field, $\xi$, this action translates to the following Langevin equation~\cite{Kamenev2023Jan} with $\langle\xi(t)\bar\xi(t')\rangle=3\gamma_{2}\delta(t-t')$,
\begin{align}\label{eq:complexlangevin}
    i(\partial_t + \gamma_1)\eta + \dot\theta = \xi,
\end{align}
along with its complex conjugate equation. The imaginary part of Eq.~\eqref{eq:complexlangevin} reads $(\partial_t+\gamma_1)\eta={\rm Im}[\xi]$, which expresses the fact that $\eta$ relaxes with rate $\gamma_{1}$ with a steady-state variance of $\langle\eta^2\rangle=3\gamma_2/(4\gamma_1)$. The real part of Eq.~\eqref{eq:complexlangevin} reads $\dot\theta={\rm Re}[\xi]$, which describes a simple Brownian motion with diffusion constant $3\gamma_2/4$. This reveals how the limit cycle is washed out by the fluctuations when $\gamma_1\sim\gamma_2$: The radial fluctuations are on the scale of the limit cycle radius, $\langle\eta^2\rangle\sim 1$, and the angular diffusion constant is comparable to the radial relaxation rate.

We note that the phase diffusion can be recast as the Fokker-Planck equation $\partial_t P(\theta,t) = \frac{3\gamma_2}{4}\partial_\theta^2P(\theta,t)$, which has eigenmodes $P_l(\theta,t) = \mathrm{e}^{il\theta-\lambda_{l}t}$ with eigenvalues $\lambda_{l}=3\gamma_2l^2/4$ for $l\in\mathbb{Z}$. These give the lowest eigenvalue branch in the Liouvillian spectrum and capture the slow angular dynamics in the limit cycle phase~\cite{dutta_quantum_2025}. 

\begin{figure}
\centering
\includegraphics[width=\linewidth]{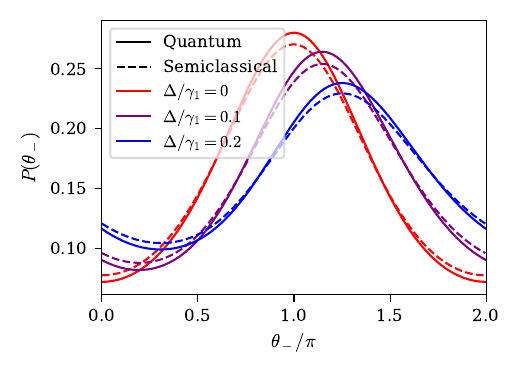}
\caption{Distribution of the phase difference $\theta_-$ at $\gamma_2/\gamma_1=0.1$ computed from the semi-classical Langevin equation and the Lindblad master equation.}
\label{fig:phasedistribution}
\end{figure}
\textit{Synchronization of two limit cycles.} We now investigate two Stuart-Landau limit-cycle oscillators which interact dissipatively through the jump operator $L_D = \sqrt{D}( \hat{a}_1+\hat{a}_2)$. Here $\hat{a}_{1,2}$ denote the annihilation operators for the two otherwise identical oscillators with detuning $\Delta=\omega_1-\omega_2$. A reactive component to the coupling can readily be included, but for concreteness we focus here on the dissipative case, which is closer to the microscopic model considered below. In this case, the photon self-energy has a matrix structure, with entries $\Pi_{11}^R=\Pi_{22}^R=(\gamma_1-D)i/2$ and $\Pi_{12}^R=\Pi_{21}^R=-iD/2$.  We follow the same procedure as above (cf. SM) to first solve two coupled saddle-point equation to determine the common synchronization frequency, $\nu=(\omega_1+\omega_2)/2$, together with two identical limit-cycle radii, $r_{1,2}^2 = (\gamma_1 - D+D\mathrm{Re}[\sqrt{1-\Delta^2/D^2}])/\gamma_2$, and then derive a Langevin equation describing the angular diffusion dynamics of $\theta_\pm = \theta_1\pm\theta_2$ as $\dot\theta_+ = \xi_+$ and
\begin{equation}\label{eq:noisyadler}
\dot\theta_- = \Delta + D \sin\theta_- + \xi_-.
\end{equation}
Here, $\xi_\pm$ denote real-valued white-noise Hubbard-\newpage
\noindent Stratonovich fields, with variances $\expval{\xi_\pm^2}=(3\gamma_1 - D+D{\rm Re}[\sqrt{1-\Delta^2/D^2}](2\mp 1))/r_n^2$. The Fokker-Planck equation corresponding to the noisy Adler equation~\eqref{eq:noisyadler} is readily solved for the steady-state distribution function, $P(\theta_-)$, expressed in terms of continued fractions~\cite{Pikovsky2001Oct}. This is shown in Fig.~\ref{fig:phasedistribution} and compared with the same quantity computed directly from the steady-state density matrix obtained by solving the corresponding LME. The small differences result from neglecting the higher order terms in $\eta$ and $\chi/r$, similar to a truncated Wigner expansion~\cite{polkovnikov_phase_2010,kato_semiclassical_2019}.

\begin{figure}
    \centering
    \includegraphics[width=\linewidth]{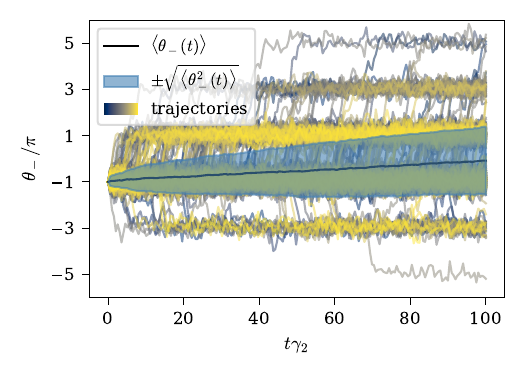}
    \caption{Trajectories of Eq.~\eqref{eq:noisyadler} at $\gamma_{2}=0.05\gamma_{1}$ (corresponding to $n=10$ photons) with $D=0.1\gamma_1$ and $\Delta=0.13D$, showing how the quantum fluctuations degrade the synchronization through phase slips. The diffusion constant $\sigma_-^2$ can be found from a linear fit of the variance (cf. Fig.~\ref{fig:effectivediffusion}).}
    \label{fig:trajectories}
\end{figure}

To assess the quality of the synchronization, we may now analyze the long-time stochastic dynamics of $\theta_{-}$. Eq.~\eqref{eq:noisyadler} describes stochastic dynamics in a tilted washboard potential for which the trajectories are shown in Fig.~\ref{fig:trajectories}. These trajectories display epochs of phase locking, $\theta_{-}\approx n\pi$ for odd-integer $n$ up to a residual noise level, interrupted by sudden phase slips, with a bias towards positive since $\Delta=0.13D$ is positive. We define the respective diffusion constants of $\theta_\pm$ as 
\begin{equation}
    \sigma_{\pm}^2 = \lim_{t\to\infty}\frac{\langle\theta_\pm(t)^2\rangle - \langle\theta_\pm(t)\rangle^2}{2t},
\end{equation}
where $\sigma_+^2=\langle\xi_+^2\rangle/2$
since $\theta_+$ follows simple Brownian motion while $\sigma_-$ may be computed  using Eq.~(22) in Ref.~\onlinecite{reimann_diffusion_2002}. The ratio of $\sigma_{-}^2$ to the noise level, $\sigma_0^2=\expval{\xi_-^2}/2$, quantifies the robustness of phase locking against noise and thereby the quality of synchronization. The smaller the ratio, the longer the epochs of perfect synchronization.

\begin{figure}
    \centering
    \includegraphics[width=\linewidth]{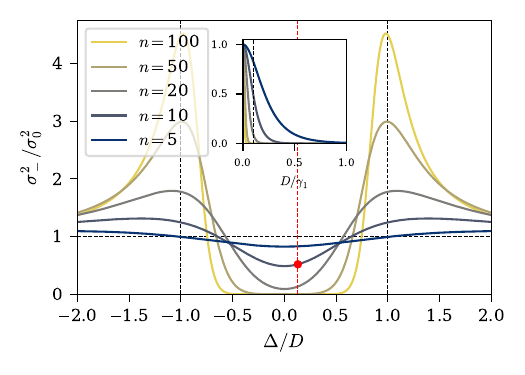}
    \caption{Effective diffusion constant of the $\theta_-$ dynamics with $D=0.1\gamma_1$ normalized by the $\eta_-$ variance $\sigma_0^2=\expval{\eta_-^2}/2$. The photon number, $n=\gamma_1/(2\gamma_2)$, is varied by changing $\gamma_2$. The red gridline and dot indicate the parameters used in Fig.~\ref{fig:trajectories}. The inset shows the diffusion constant at $\Delta=0$ vs $D$ with a gridline at $D=0.1\gamma_{1}$.}
    \label{fig:effectivediffusion}
\end{figure}

In Fig.~\ref{fig:effectivediffusion}, this ratio is plotted against detuning for a weak dissipative coupling $D=0.1\gamma_1$ at different photon numbers, $n=\langle \hat{a}_{m}^{\dagger}\hat{a}_{m}\rangle\vert_{D=0}=r_{m}^2/2=\gamma_{1}/(2\gamma_{2})$, at zero coupling. The black vertical gridlines in Fig.~\ref{fig:effectivediffusion} demarcate the synchronization region, $\abs{\Delta}/D<1$, of the deterministic Adler equation corresponding to Eq.~\eqref{eq:noisyadler} without noise. At this boundary, the system exhibits a pronounced maximum in the diffusion constant. For a 5-photon limit cycle, $\sigma_-^2/\sigma_0^2$ is close to unity, even at vanishing detuning, indicating that there is no real improvement in the phase locking due to the coupling: A nearly complete lack of synchronization. As shown in the inset,  the phase diffusion rate vanishes as $D\approx \gamma_1$ for all radii of the limit-cycles. Nevertheless, this is due to a trivial mode coupling of the two oscillators and not to synchronization. 

Earlier studies of synchronization between two quantum Stuart-Landau oscillators or entrainment of a single oscillator to an external drive~\cite{lee_quantum_2013, walter_quantum_2014, walter_quantum_2015, sonar_squeezing_2018, Mok2020Sep, hermansen_synchronizing_2026,nadolny2025quantumlimitcyclessynchronization} have primarily used the angular distribution, $P(\theta_{-})$, and observed detuning as a measure of synchronization. Nevertheless, these are both mod-$2\pi$ measures of so-called {\it imperfect synchronization} (a terminology used for classical chaotic and stochastic dynamical systems) which are insensitive to the long-time diffusion rate of $\theta_{-}$~\cite{Zaks1999May,Boccaletti2002Aug,freund2003}. 

For few-photon limit cycles the radial dynamics becomes increasingly important, and the phase variable of the limit cycle loses its meaning. In the Stuart-Landau example studied here, already the 5-photon trajectories do not provide for a clear separation into radial and angular dynamics (cf.~SM). Our analysis is therefore meaningless in the deep quantum limit ($\gamma_2/\gamma_1\gg 1$)~\cite{walter_quantum_2014, walter_quantum_2015, sonar_squeezing_2018, Mok2020Sep} since the notion of limit cycles, and thereby of their synchronization, is lost altogether. In this limit, the two-photon loss effectively reduces the Fock state space to $n=0,1$, and the frequency locking observed fx. in Ref.~\cite{walter_quantum_2015} rather corresponds to a level attraction, due to a strong dissipative mode coupling.

\textit{Non-Markovian effects.} The above analysis can be extended to the more general case in which the photon self-energy matrix for the two oscillators,  $\Pi_{mn}^{\alpha}(\omega)$ with $\alpha=R,A,K$, depends on frequency. We restrict the analysis to the case with identical self-energies, $\Pi_{11}^{\alpha}(\omega)=\Pi_{22}^{\alpha}(\omega)$, and with symmetric (generally both reactive and dissipative, but predominantly dissipative for the parameters used below) mode couplings $\Pi_{12}^{\alpha}(\omega)=\Pi_{21}^{\alpha}(\omega)$. As we show in the SM, the corresponding saddle-point equations comprise a set of coupled non-linear implicit differential equations which are hard to solve in general. Nevertheless, focusing our attention on synchronized saddle-point solutions, the problem reduces to solving a far simpler algebraic equation for their common synchronized frequency, $\nu$, along with the radii $r_{1,2}$ and the static phase difference, $\theta_0$, given (with $\sigma_1 =-\sigma_2=1$) as
\begin{equation}\label{eq:sync_algebraiceq}
    \nu-\omega_j - \Pi^R_{11}(\nu) - \mathrm{e}^{i\sigma_j\theta_0} \Pi^R_{12}(\nu) \qty(\frac{r_2}{r_1})^{\sigma_i} = \Lambda_1^* r_i^2,
\end{equation}

To analyze the breakdown of synchronization due to quantum noise in this non-Markovian case, one may determine the effective action for the fluctuations around the limit cycles, as shown in detail in the SM. 
Assuming $\Pi_{mn}^{\alpha}(\omega)$ to be slowly varying, we expand the retarded self-energy as $\Pi_{mn}^R(\nu+i\partial_t) \approx \Pi^R_{mn}(\nu) + (\partial_{\omega}\Pi^R_{mn})i\partial_t$, and evaluate the Keldysh self-energy at the synchronization frequency, $\nu$. This leads to the following It$\hat{\rm o}$-type stochastic equation
\begin{align}\label{eq:nonmarkovlangevin}
    \sum_{n}\!A_{mn}\dot\phi_n \!= r_m^2\Lambda_1^*\bar\phi_m\phi_m^2+\!\sum_n\!\qty[\Pi^R_{mn}(\nu)\phi_n \!+ \!B_{mn}\xi_n],
\end{align}
where $\phi_n=\mathrm{e}^{-i\theta_n}(1+\eta_n)$ are the rescaled classical fields and $A_{mn}=i\delta_{mn}-\partial_{\omega}\Pi^R_{mn}(\nu)$ and $B_{mn}$ describe respectively the friction and multiplicative noise correlations (cf. SM). The complex Hubbard-Stratonovich fields, $\xi_n$, are normalized to have correlations $\langle\bar\xi_m(t)\xi_n(t')\rangle=\delta_{mn}\delta(t-t')$. Solving Eq.~\eqref{eq:nonmarkovlangevin} numerically, one finds that $\int\dd t\langle \bar\phi_m(t)\phi_n(t+\tau)\rangle\sim\mathrm{e}^{-\tau\gamma_2}$, implying that the adiabatic approximation used above is justified when $\Pi^{\alpha}_{mn}(\omega)$ varies only very little over a frequency range of $\gamma_{2}$~\cite{hriscu_quantum_2013}.   

\begin{figure}
    \centering    \includegraphics[width=\linewidth]{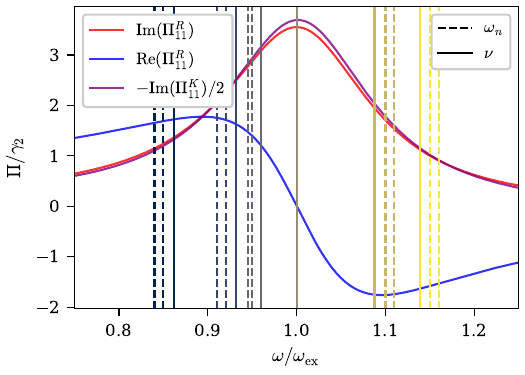}
    \caption{Self-energies $\Pi^{R,K}_{11}(\omega)=\Pi^{R,K}_{22}(\omega)\approx -\Pi^{R,K}_{12}(\omega)=-\Pi^{R,K}_{21}(\omega)$ for the oscillators coupled via the DQD. The vertical dashed lines indicate the bare resonator frequencies $\omega_1$ and $\omega_2$, and the solid lines mark the corresponding synchronization frequency $\nu$ determined from Eq.~\eqref{eq:sync_algebraiceq}.}
    \label{fig:piR}
\end{figure}

As a concrete example, we consider a gain medium comprised by a voltage-biased double quantum dot (DQD), which has been demonstrated experimentally to provide for limit-cycle dynamics in a superconducting microwave resonator~\cite{liu_semiconductor_2015}, and predicted theoretically to give rise to synchronization, when capacitively coupling a resonator to each QD~\cite{hermansen_synchronizing_2026}. The photon self-energy matrix, $\Pi^{\alpha}_{mn}(\omega)$, encodes the electrical polarizability of the driven-dissipative DQD system, which resembles a simple Lorentz model for the permittivity. For a bias voltage larger than the electronic excitation energy of the DQD, $\hbar\omega_{\rm ex}$, the imaginary part of the retarded photon self-energy has an approximately Lorentzian peak centered at $\omega_{\rm ex}$, the width of which is set by the electron tunneling rate to the metallic leads, $\Gamma$. In this microscopic model, the nonlinearity has been calculated perturbatively~\cite{hermansen_synchronizing_2026,cottet_theory_2020}, but for simplicity, we use here the same non-linearity~\eqref{eq:s4} as above with $\gamma_{2}=\Gamma/10$, consistent with the adiabatic approximation and a weak electron-photon coupling. Using nearly the same parameters~\footnote{In the notation of Ref.~\onlinecite{hermansen_synchronizing_2026}, we choose the DQD parameters as $t=\varepsilon=\omega_{\rm ex}/\sqrt{8}$ and $\Gamma=0.05\omega_{\rm ex}$, while the electron-photon coupling is set to $(g/\omega_{\rm ex})^2 = 0.02$ and the bias voltage at $eV=10\hbar\omega_{\rm ex}$.} as in Ref.~\onlinecite{hermansen_synchronizing_2026}, we show the self-energies $\Pi^\alpha_{11}(\omega)$ in Fig.~\ref{fig:piR}, together with the synchronization frequency $\nu$ found from solving the saddle-point equations for different values of the bare oscillator frequencies. Interestingly, the synchronization frequency is no longer simply the mean of the bare frequencies. Instead, it is shifted towards the excitation frequency, $\omega_{\rm ex}$, where the gain ($\rm{Im}[\Pi^R(\omega)]$) attains its maximum. 

\begin{figure}
    \centering
    \includegraphics[width=\linewidth]{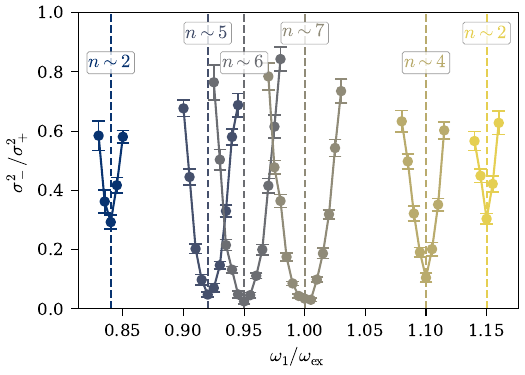}
    \caption{Ratios of angular diffusion constants plotted against $\omega_1$ for the six different values, $\omega_2/\omega_{\rm ex}=0.84, 0.92,0.95,1.00,1.10,1.15$, used also in Fig.~\ref{fig:piR}. The dashed lines indicate the values of $\omega_2$ and the approximate photon numbers are indicated for the limit cycles with $\omega_1=\omega_2$. Only the shown points provide a stable synchronized solution to the saddle-point equation~\eqref{eq:sync_algebraiceq}.}
    \label{fig:sigmaratio}
\end{figure}

Integrating Eq.~\eqref{eq:nonmarkovlangevin} over many realizations of the Hubbard-Stratonovich fields, we compute the diffusion constant, $\sigma_\pm$, from a linear fit of the variances $\langle\theta_\pm^2\rangle(t)$ as a function of time. 
In Fig.~\ref{fig:sigmaratio}, we show the ratio of diffusion constants, $\sigma_-^2/\sigma_+^2$, as a function of $\omega_1$ for different values of $\omega_2$. As in the Markovian case, $\theta_+$ essentially undergoes Brownian motion, meaning that $\sigma_+^2$, quantifies the bare noise level, while $\sigma_-^2$ is suppressed when $\omega_1\approx \omega_2$ due to the phase locking. The minimum value of $\sigma_-^2/\sigma_+^2$, corresponding to the best synchronization, is attained for $\omega_1=\omega_2\approx\omega_{\rm ex}$. However, both the extent of the synchronized region of detuning (Arnold tongue), set by the widths, and the quality of synchronization itself at $\omega_1=\omega_2$, set by the minima of the parabolas in Fig.~\ref{fig:sigmaratio}, generally depend on the location of the synchronization frequency $\nu$ compared to $\omega_{\rm ex}$. 
At the outermost frequencies ($\omega_{2}/\omega_{\rm ex}=0.85, 1.12$), $\sigma_-^2/\sigma_+^2$ is larger, indicating markedly shorter epochs of perfect synchronization. Furthermore, since the gain is smaller at these off-resonant frequencies, the limit cycles are practically washed out by the noise (cf. SM). At $\omega_1=\omega_2\approx\omega_{\rm ex}$ the limit cycles are robust and the epochs of phase locking are longer. However, closer inspection shows that the effective ratio of coupling to gain, as dictated by this specific microscopic model, corresponds to having $\sim D/\gamma_{1}\approx 0.5$ in the Stuart-Landau model (cf. inset of Fig.~\ref{fig:effectivediffusion}), which already challenges the assumption of weakly coupled systems distinguishing synchronization from mode coupling~\cite{Pikovsky2001Oct}. 

\textit{Summary and Outlook.} We have analyzed the synchronization of continuous variable limit-cycle oscillators in the presence of quantum fluctuations. The proliferation of quantum phase slips degrades the synchronization to an imperfect (mod-$2\pi$) synchronization characterized by epochs of phase locking interrupted by sudden phase slips. The duration of these epochs, set by the inverse of the phase-difference diffusion constant, provides a measure for the quality of synchronization. This measure allows to distinguish genuine phase-locking from the more trivial dissipative mode coupling, which does not require a limit cycle. 

To illustrate this quantum desynchronization mechanism, we have analyzed the phase dynamics for the paradigmatic model of two coupled Stuart-Landau oscillators, as well as for a fully microscopic and non-Markovian model of two microwave resonators coupled via a voltage-biased DQD. In both cases we find that the quantum fluctuations reduce the quality of synchronization, while frequency, and phase correlations remain. 
The non-Markovian case with a resonant gain medium, was shown to reveal a secondary entrainment of the synchronization frequency towards the frequency for which the environment provides the largest gain. It would be interesting to explore further this self-sustaining entrainment for gain media, which are more sharply peaked. This requires the inclusion of higher order terms in the gradient expansion for the photon self-energy used here and is relegated to future work.



\textit{Acknowledgements.} We thank Fabian Hassler, Mathias Heltberg, Mogens H. Jensen, Steven Kim, and Christopher W. Wächtler for useful discussions. H.~C. acknowledges support from the Novo Nordisk Foundation grant NNF20OC0060019.

\bibliography{refs}

\end{document}


\title{Quantum Desynchronization of Limit Cycles: Supplementary Material}

\author{Hans Christiansen}
\author{Jens Paaske}
\affiliation{Center for Quantum Devices, Niels Bohr Institute, University of Copenhagen, 2100 Copenhagen, Denmark}

\maketitle

\section{Derivation of the Noisy Adler Equation}

In this section, we give the details on the derivation of Eq.~(11) in the main text. First we describe the transformation from the Lindblad Master equation (LME) to a coherent-state Keldysh path integral, in the case of two Stuart-Landau oscillators with a dissipative interaction. In general, the self-energies for a set of bosonic modes described by annihilation operators $\hat a_i$, that interact coherently through the Hamiltonian $H=\sum_{ij}\hat{a}_i \Delta_{ij}\hat{a}_j$ and jump operators $\hat L_{+,v} = \sum_{j}\nu_{vj}\hat{a}_j^\dag$ and $\hat L_{-,v}=\sum_{j}\mu_{vj}\hat{a}_j$ are given by~\cite{thompson_field_2023}
\begin{align}
    \Pi^R = \Delta - \frac{i}{2}(\mu^\dag \mu - \nu^\dag \nu ), && \Pi^K = -i (\mu^\dag \mu + \nu^\dag \nu ) . 
\end{align}
In particular, the LME for two Stuart-Landau oscillators with a dissipative interaction is given by~\cite{walter_quantum_2015}
\begin{equation}
    \partial_t\hat\rho(t) = - i\sum_n \big[\omega_n \hat{a}_n^\dag \hat{a}_n,\hat\rho(t)\big]+\sum_j \qty(\hat{L}_j\hat\rho(t) \hat{L}_j^\dag-\frac{1}{2}\{\hat{L}_j^\dag \hat{L}_j,\hat\rho(t)\}),
\end{equation}
with $\hat L_j =\{\sqrt{\gamma_1}\hat a_1^\dag,\sqrt{\gamma_1} \hat a_2^\dag,\sqrt{D}(\hat a_1+\hat a_2),\sqrt{\gamma_2}\hat a_1^2,\sqrt{\gamma_2}\hat a_2^2\}$. The linear jump operators are equivalent to the frequency independent self-energies 
\begin{align}\label{eq:slselfenergy}
    \Pi^R = -\frac{i}{2}\mqty( D-\gamma_1 && D \\ D && D-\gamma_1), && \Pi^K = -i\mqty(D+\gamma_1 && D \\ D && D+\gamma_1).
\end{align}
Expressing the quantum and classical fields of each oscillator, $n=1,2$, as 
\begin{equation}
\mqty(\phi^{\rm cl}_n\\\phi^{\rm q}_n) = \mathrm{e}^{-i\theta_n - i\nu_n t}\mqty(r_m(1+\eta_n) \\  \chi_n/r_n) .
\end{equation}
the action consisting of the self-energies in Eq.~\eqref{eq:slselfenergy} and the quartic non-linearity coming from the two-photon loss, $\sqrt{\gamma_2}\hat a_n^2$, is given as 
\begin{align}\label{eq:action_multipleoscilators_markov}
    S&=\sum_{m,n}\int\dd t \bigg[\mqty( 1+\eta_m \\ \bar\chi_m)^{T}\mqty( 0 & P^A_{mn} \\  P^R_{mn}& P^K_{mn}/(r_n r_m))
    \mqty( 1+\eta_n \\ \chi_n )  \bigg]-\sum_n\int dt\qty[\Lambda_5 \bar\chi_n\chi_n + r_n^2(1+3\eta_n) (\Lambda_1^*\bar\chi_n+\Lambda_1\chi_n)],
\end{align}
with $\Lambda_1^*=-i\gamma_2/2$ and $\Lambda_5=-2i\gamma_2$ and where $m,n=1,2$. The inverse propagators are given by 
\begin{align}
    &P^R_{mn} = (\nu_m +i\partial_t+\dot\theta_m-\omega_m)\delta_{mn} - \mathrm{e}^{i(\theta_m-\theta_n)}\Pi_{mn}^R\frac{r_n}{r_m}, && P^K_{mn} = - \mathrm{e}^{i(\theta_m-\theta_n)}\Pi_{mn}^K/(r_nr_m).
\end{align}
  The saddle-point equations, $\delta S/\delta \bar\chi_n\vert=0$, read
\begin{align}
    (\dot\theta_m + \nu_m -\omega_m)  - \sum_{n}\mathrm{e}^{i(\theta_m-\theta_n)+t(\nu_m-\nu_n)}\qty(\frac{r_n}{r_m})\Pi_{mn}^R=\Lambda_1^* r_m ^2.
\end{align}
A synchronized solution requires that $\dot\theta_m=0$ and $\nu_1=\nu_2= \nu$, in which case the equations take the following form 
\begin{equation}
    \nu -\omega_m - \Pi_{11}^R-\mathrm{e}^{i\sigma_m\theta_0}\Pi_{12}^R\qty(\frac{r_2}{r_1})^{\sigma_m}= r_m^2\Lambda_1^*,
\end{equation}
where $\theta_0 = \theta_1-\theta_2$ is the static phase difference and we use $\sigma_1=-\sigma_2=1$. In this case the solution is given by $\nu=(\omega_1+\omega_2)/2$ and
\begin{align}
    \sin \theta_0=- \frac{\Delta}{D}, && r_n^2 =\frac{\gamma_1 - D\qty(1-\mathrm{Re}\sqrt{1-\Delta^2/D^2})}{\gamma_2},
\end{align}
and these solutions are stable inside the Arnold tongue, $\abs{\Delta}< D$. When $\abs{\Delta}>D$, $\nu_1\neq\nu_2$ so the coupling term is fast oscillating. Neglecting this fast oscillating term, the saddle-point equations essentially decouple and the solution becomes $\nu_n=\omega_n$ and $r_n^2=(\gamma_1-D)/\gamma_2$. 

Inserting the saddle-point values into Eq.~\eqref{eq:action_multipleoscilators_markov} and expanding to second order in $\dot \theta_n$, $\eta_n$, $\bar\chi_n$ and $\chi_n$ we obtain
\begin{align}\label{eq:action_multipleoscilators2}
    &S[\theta_n,\eta_n,\bar\chi_n,\chi_n] = \sum_{n\in\{1,2\}}\int \dd t\bigg[\bar\chi_n\qty(\dot\theta_n-\mathrm{e}^{i\sigma_n(\theta_1-\theta_2)}\Pi_{12}^R+\mathrm{e}^{i\sigma_n\theta_0}\Pi_{12}^R)+\qty(\dot\theta_n-\mathrm{e}^{-i\sigma_n(\theta_1-\theta_2)}\Pi_{12}^A+\mathrm{e}^{-i\sigma_n\theta_0}\Pi_{12}^A)\chi_n \bigg]\notag\\& \sum_{mn}\int\dd t\mqty(\eta_m\\\bar\chi_m)\mqty(0 && (\nu-\omega_m +i\partial_t)\delta_{mn}-\mathrm{e}^{-i(\theta_m-\theta_n)}\Pi^A_{mn}-3\delta_{mn}r_m^2\Lambda_1 \\ (\nu-\omega_m +i\partial_t)\delta_{mn}-\mathrm{e}^{i(\theta_m-\theta_n)}\Pi^R_{mn}-3\delta_{mn}r_m^2\Lambda_1^* && P^K_{mn}-\Lambda_5\delta_{mn})\mqty(\eta_n\\\chi_n),
\end{align} 
where the Keldysh kernel depends on the $\theta_n$ fields leading to a multiplicative noise. However, if we set the phase difference equal to its saddle-point value $\theta_1-\theta_2\to\theta_0$, we can write the term quadratic in the quantum field as $\sum_{mn}\bar\chi_miC_{mn}\bar\chi_n$ with the correlation matrix
\begin{align}
    C &= \frac{1}{r^2}\mqty( 3\gamma_1 - 2D\cos\theta_0-D && D\mathrm{e}^{i\theta_0} \\ D\mathrm{e}^{-i\theta_0} && 3\gamma_1 - 2D\cos\theta_0-D).
\end{align}
We decouple the quadratic $\bar\chi_m\chi_n$ terms via complex Hubbard-Stratonovich fields $\xi_1$ and $\xi_2$ using the identity 
\begin{equation}
    \exp{-\int \dd t \bar \chi_m C_{mn} \chi_n}=\prod_m\int\mathcal D[\bar\xi_m,\xi_m]\exp{-\sum_{mn}\int \dd t \bar \xi_m [C^{-1}]_{mn} \xi_n - i\sum_{mn}\int \dd t \qty(\xi_m \chi_n +  \bar\chi_m\xi_n)}.
\end{equation}
Inserting this into the partition function, $Z=\int \prod_nD[\theta_n,\eta_n]D[\bar\chi_n,\chi_n]\exp{iS}$ leaves only terms linear in $\bar\chi_n$ and $\chi_n$. This allows us to integrate out the quantum field leaving functional $\delta$-functions in the partition function that forces the classical fields $\theta_n$ and $\eta_n$ to obey the Langevin equation 
\begin{equation}
    \qty(-\sigma_m\frac{\Delta}{2} + i\partial_t-\Pi_{11}^R-3r_m^2\Lambda_1^*)\eta_m -\mathrm{e}^{i\sigma_m(\theta_1-\theta_2)}\Pi_{12}^R\eta_{\bar m} + \dot\theta_m-\mathrm{e}^{i\sigma_m(\theta_1-\theta_2)}\Pi_{12}^R+\mathrm{e}^{i\sigma_m\theta_0}\Pi_{12}^R = \xi_m
\end{equation}
with $\langle \bar\xi_m\xi_n\rangle = C_{mn}$ and $\bar 0 = 1$, $\bar 1 = 0$. Inserting the expressions for the self energies, the real part becomes 
\begin{equation}
    \dot\theta_m - \sigma_m \frac{\Delta}{2}(1+\eta_m) - \frac{D\sigma_m}{2}\sin(\theta_1-\theta_2)(1+\eta_{\bar m}) = \xi_m',
\end{equation}
where $\xi_n'$ is the real part of the HS fields. As for the single oscillator case, the $\eta_n$ field fluctuations are on the scale $\langle \eta_n^2\rangle \sim \gamma_2/\gamma_1\ll 1$. Neglecting these fluctuations, we arrive at two decoupled equations for the sum and difference phases, $\theta_\pm = \theta_1\pm\theta_2$, that read
\begin{align}\label{eq:SMadler}
    \dot\theta_- = \Delta + D  \sin\theta_- + \xi'_-, &&
    \partial_t\theta_+ = \xi'_+, 
\end{align}
where $\xi'_\pm = \xi_{1}'\pm\xi_2'$ are the real parts of the complex fields $\xi_\pm$ with $\langle \bar\xi_\pm(t)\xi_\pm(t')\rangle = 2\langle\xi'_\pm(t)\xi'_\pm(t')\rangle$
where
\begin{equation}
    \langle\xi'_\pm(t)\xi'_\pm(t')\rangle=\delta(t-t')[C_{11}+C_{22}\pm(C_{12}+C_{21})]/2 = \delta(t-t')\qty(3\gamma_1 - D+D\mathrm{Re}\qty[\sqrt{1-\Delta^2/D^2}](2\mp 1))/r_n^2,
\end{equation}
sets the noise fluctuations in Eq.~\eqref{eq:SMadler}. In the main text we suppress the primes on $\xi_\pm'$ indicating a real-valued field.

\section{Langevin Equations for non-Markovian Bath}

\begin{figure}
    \centering
    \includegraphics[width=\linewidth]{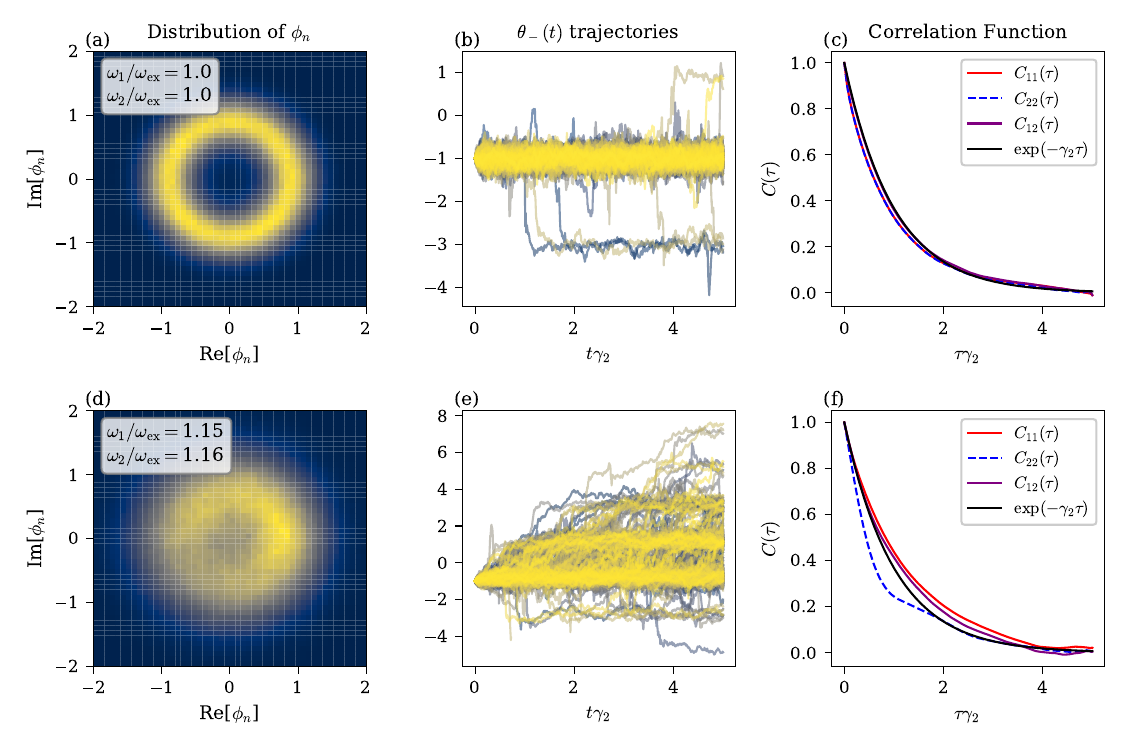}
    \caption{(a,d) distribution of the complex trajectories, $\phi_{1,2}(t)$, (b,e) $\theta_-(t)=\theta_1(t)-\theta_2(t)$  trajectories, (c,f) autocorrelation function $C_{mn}(\tau)$ showing that the correlation time is given roughly as $\gamma_2$. Parameters are the same as in the main text with $\omega_1=\omega_2=\omega_{\rm ex}$ in (a,b,c) and $\omega_1=1.15\omega_{\rm ex}$ and $\omega_2=1.16\omega_{\rm ex}$ in (d,e,f).}
    \label{fig:correlationfunction}
\end{figure}

\begin{figure}
    \centering
    \includegraphics[width=\linewidth]{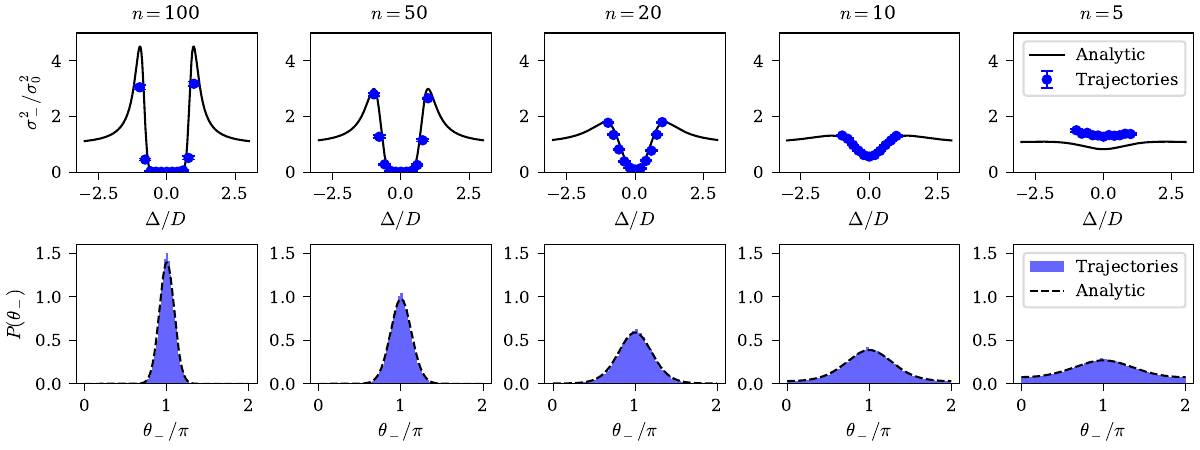}
    \caption{Phase-difference diffusion constant $\sigma_-^2=\lim_{t\to\infty}\langle\theta(t)^2\rangle/2t$ and distribution $P(\theta_-)$ for the two coupled Stuart-Landau oscillators, computed from the trajectories or the analytic model discussed in the main text. For the phase diffusion, the analytic prediction agrees with the trajectories in all cases except at $n=5$ photons, where the radial dynamics creates even more phase slips, which the analytic model neglects.}
    \label{fig:sl_diffusion_phasedist}
\end{figure}

Here we show the derivation of the saddle-point equations, effective action and resulting Langevin equations for two limit cycles coupled through a non-Markovian gain medium. As in the main text, we rescale the fields as
\begin{equation}
\mqty(\phi^{\rm cl}_n\\\phi^{\rm q}_n) = \mathrm{e}^{- i\nu_n t}\mqty(r_n\phi_n \\  \chi_n/r_n) 
\end{equation}
Written in these fields, the action for the two coupled oscillators ($m,n\in\{1,2\}$) becomes
\begin{align}\label{eq:action_multipleoscilators}
    S&=\sum_{m,n}\int\dd t\mqty( \bar\phi_m \\ \bar\chi_m)^{T}\mqty( 0 & P^A_{mn}(\nu_m+i\partial_t) \\  P^R_{mn}(\nu_m+i\partial_t)& P^K_{mn}(\nu_m+i\partial_t))
    \mqty( \phi_n \\ \chi_n ) -\sum_n\int dt\qty[\Lambda_5 \bar\chi_n\chi_n\bar\phi_n\phi_n +r_n^2\Lambda_1^*\bar\chi_n\bar\phi_n\phi_n^2+r_n^2\Lambda_1\bar\phi_n\phi_n^2\chi_n],
\end{align}
with inverse propagators given by
\begin{align}
P^R_{mn}(\omega) = (\omega-\omega_n)\delta_{mn} - \Pi_{mn}^R(\omega)r_n/r_m, &&
    &P^K_{mn}(\omega) = - \Pi_{mn}^K(\omega)/(r_n r_m),
\end{align}
where the oscillators are assumed to be identical except for their bare frequencies,  $\omega_{1,2}$. The saddle point equations, $\delta S/\delta \bar\chi_n=0$, can be written as
\begin{align}
    \sum_n P_{mn}^R(\nu_n+i\partial_t)\phi_n - \Lambda^*_1r_m^2\bar\phi_m\phi_m^2 = 0,
\end{align}
due to the derivative operator in the self-energy these are implicit differential equations that are generally hard to solve. However a synchronized solution requires that $\dot\phi_1=\dot\phi_2 =0$ and $\nu_1=\nu_2= \nu$. Since we absorb the static scaling of the $\phi_n$ fields into $r_n$, we can set $\abs{\phi_n}=1$ and write them as $\phi_n=\mathrm{e}^{-i\theta_n}$. Denoting the static phase difference by $\theta_0=\theta_1-\theta_2$, the saddle-point equations become non-linear \textit{algebraic} equations (using $\sigma_1=-\sigma_2=1$)
\begin{align}
\nu-\omega_n - \Pi_{11}^R(\nu) - \mathrm{e}^{i\sigma_n\theta_0}\Pi_{12}(\nu)\qty(\frac{r_2}{r_1})^{\sigma_n} = \Lambda_1^*r^2_n
\end{align}
where $\sigma_1=-\sigma_2=1$. This equation can be solved for the 4 real parameters $\{\nu,\theta_0,r_1,r_2\}$. Inserting these saddle-point solutions into the nonlinear part of the action, we obtain:
\begin{align}
    S&=\sum_{m,n}\int\dd t\mqty( \bar\phi_m \\ \bar\chi_m)^{T}\mqty( 0 && P^A_{mn}(\nu+i\partial_t)-r_n^2\Lambda_1\bar\phi_m\phi_n\delta_{nm} \\  P^R_{mn}(\nu+i\partial_t)-r_n^2\Lambda_1^*\bar\phi_m\phi_n\delta_{nm} && P^K_{mn}(\nu+i\partial_t)-\Lambda_5\bar\phi_m\phi_n\delta_{mn})
    \mqty( \phi_n \\ \chi_n )
\end{align}
Approximating $P^K_{mn}(\nu+i\partial_t) \approx P^K_{mn}(\nu)$ and defining $C_{mn}=-i\qty(P_{mn}^K(\nu)-\Lambda_5\bar\phi_m\phi_n\delta_{mn})$, we use the identity 
\begin{equation}
    \exp{-\int \dd t \bar \chi_m C_{mn} \chi_n}= \int\mathcal D[\bar\xi,\xi]\exp{-\sum_n\int \dd t \bar \xi_n \xi_n - i\sum_{mn}\int \dd t \qty(\xi_m B_{mn}^T\chi_n +  \bar\chi_m B_{mn}\xi_n)},
\end{equation}
where $B$ is given as 
\begin{equation}
    B = \mqty( \sqrt{C_{11}} & 0 \\ C_{12}/{\sqrt{C_{11}}} & \sqrt{C_{22}-C_{12}^2/C_{11}}),
\end{equation}
such that $C=BB^T$. This allows us to integrate over the quantum fields to recover a Langevin equation. to solve this, we expand the retarded self-energy to first order as $\Pi^R_{nm}(\nu+i\partial_t) \approx \Pi_{nm}^R(\nu) + \partial_{\omega}\Pi^R_{mn}(\nu)i\partial_t$, so we can write the retarded Langevin equation as 
\begin{equation}
    \sum_nA_{mn}\dot\phi_n = \sum_n\Pi^R_{mn}(\nu)\phi_n + r_n^2\Lambda_1^*\bar\phi_n\phi_n^2 + \sum_n B_{mn}\xi_n.
\end{equation}
with $A_{mn} =i\delta_{mn}-\partial_{\omega}\Pi^R_{mn}(\nu)$. Using the same model as given in the main text, in Fig.~\ref{fig:correlationfunction}, we show the correlation functions
\begin{equation}
    C_{mn}(\tau) =\frac{1}{T}\int_0^T\dd t\langle\bar\phi_m(t)\phi_n(t+\tau)\rangle,
\end{equation}
and find that the exponential correlation time is roughly given as the inverse of $\gamma_2$. This slow dynamics is what justifies the adiabatic approximation used above. Also shown are the corresponding trajectories $\theta_-(t)$ and histograms showing the distribution of $\phi_1$ and $\phi_2$. We can now compute the variance of $\theta_\pm(t)$ as a function of time to compute the diffusion constants shown in the main text. In Fig.~\ref{fig:sl_diffusion_phasedist}, we compare the diffusion constants computed from the complex Langevin equation against the noisy Adler equation derived in the previous section, for the coupled Stuart-Landau oscillators. Here we find good agreement between the two approaches except at $n=5$ photons, where the radial dynamics contributes significantly to the resulting diffusion. 

\bibliography{refs}